\documentclass[12pt,preprint]{aastex}
\begin{document}

\title{The Detection of Diffuse Extended Structure in 3C~273: Implications for Jet Power}

\author{Brian Punsly\altaffilmark{1} and Preeti Kharb\altaffilmark{2, 3}}
\altaffiltext{1}{1415 Granvia Altamira, Palos Verdes Estates CA, USA
90274 and ICRANet, Piazza della Repubblica 10 Pescara 65100, Italy,
brian.punsly@cox.net} \altaffiltext{2}{NCRA-TIFR, Post Bag 3,
Ganeshkhind, Pune 411007, India}
 \altaffiltext{3}{Indian Institute of Astrophysics, II Block, Koramangala, Bangalore 560034, India}

\begin{abstract}
We present deep Very Large Array imaging of 3C~273 in order to
determine the diffuse, large scale radio structure of this famous
radio-loud quasar. Diffuse extended structure (radio lobes) is
detected for the first time in these observations as a consequence
of high dynamic range in the 327.5 and 1365 MHz images. This
emission is used to estimate a time averaged jet power, $7.2 \times
10^{43} \rm{~ergs~s^{-1}} < \overline{Q} < 3.7 \times 10^{44}
\rm{~ergs~s^{-1}}$. Brightness temperature arguments indicate consistent
values of the time variability Doppler factor and the compactness
Doppler factor for the inner jet, $\delta \gtrsim 10$. Thus, the
large apparent broadband bolometric luminosity of the jet, $\sim 3
\times 10^{46}\rm{~ergs~s^{-1}}$, corresponds to a modest intrinsic
luminosity $\gtrsim 10^{42}\rm{~ergs~s^{-1}}$, or $\sim 1\%$ of
$\overline{Q}$. In summary, we find that 3C~273 is actually a
``typical" radio loud quasar contrary to suggestions in the
literature. The modest $\overline{Q}$ is near the peak of the
luminosity distribution for radio loud quasars and it is consistent
with the current rate of dissipation emitted from millimeter
wavelengths to gamma rays. The extreme core-jet morphology is an
illusion from a near pole-on line of sight to a highly relativistic
jet that produces a Doppler enhanced glow that previously swamped
the lobe emission.  3C~273 apparently has the intrinsic kpc scale
morphology of a classical double radio source, but it is distorted
by an extreme Doppler aberration.

\end{abstract}

\keywords{Black hole physics --- magnetohydrodynamics (MHD) ---
galaxies: jets---galaxies: active --- accretion, accretion disks}

\section{Introduction}
3C~273 is the nearest and brightest quasar in virtually all
wavebands from radio to gamma rays. It is the prototypical
quasi-stellar object and flat (radio) spectrum, core dominated
radio-loud quasar. In the standard model of quasar unification, the
flat spectrum, core dominated quasars are drawn from the same parent
population as the lobe dominated, radio-loud quasars, but only
appear core dominated due to large Doppler boosting of a highly
relativistic jet as a consequence of a nearly pole-on line of sight
towards the observer \citep{ant93}. In an effort to test the unified
scheme, deep observations were performed on representatives from a
large sample of compact radio sources as determined by snap shot
Very Large Array (VLA) observations \citep{per82}. The expectation
was that the radio lobe on the side of the quasar in which the jet
pointed towards Earth would be viewed end on and appear as a diffuse
halo surrounding a bright nuclear unresolved core. The two deepest
set of observations involved the VLA at 1.4 GHz \citep{ant85,mur93}.
Surprisingly, the core-halo configurations were less common than
expected. However, offset lobes on both the jet and counter jet
sides of the nucleus or just a single offset lobe on the jetted side
were often detected. There were a large number of cores with a one
sided jet and no lobes, core-jet configurations, and still some
naked cores. The core-jet and naked core configurations were the
most curious since it was unclear how many of these quasars had
diffuse emission that required a dynamic range beyond the capability
of the observations in order to be detected. One of the core-jet
objects was 3C~273, despite a small easterly extension half way down
the jet (see our Figure 1) that was conjectured to be evidence of a
radio lobe, however no diffuse emission was detected
\citep{dav85,con93}. It was proposed that 3C~273 might be
intrinsically one-sided \citep{dav85}. The jet structure of 3C~273
has been characterized as a ``nose cone"; such jets have been
suggested to be magnetically confined \citep{kom99}. 3C~273 has also
been referred to as a ``naked jet", that is one without a
surrounding radio lobe \citep{cla86}. One explanation for naked jet
sources is that they are young. The lobes have not formed yet since
there is insufficient time for the slow back flow of plasma from the
hot spot to fill a cocoon \citep{liu92}. However, the jet in 3C~273
is more than 150,000 light years long as projected on the sky plane
and is viewed nearly pole-on. Thus, the hot spot at the end of the
jet is $\sim 10^{6}(c/v_{\rm{adv}})$ years old, where
$v_{\rm{adv}}$ is the hot spot advance speed. 3C~273 is not a
young radio source. This raises the questions, where is the lobe
emission, or why is there no lobe emission?
\par We have pursued a program of retrieving archival observations that is based on the hypothesis that due to the extremely
bright radio core and jet, the large scale morphology has exceeded
the dynamic range limits of previous observations and for this
reason its true nature has eluded astronomers. Detection of the lobe
emission in this proto-typical quasar is fundamentally important
since the Doppler enhancement is so intense that it is difficult to
extricate any type of isotropic flux that could be used to estimate
jet power without the enormous ambiguity imposed by Doppler beaming.
Our L-band (1365 MHz) and P-band (327.5 MHz) imaging are chosen to
detect the halo emission and estimate its spectral index.
\par  The paper is organized as follows. Section
2 describes the details of the observations.  The results of
Sections 2 are used to estimate the long term time average jet power
in Section 3. In Section 4, we estimate the Doppler factor at the
base of the jet by various means and find that it is consistently
determined as $\gtrsim 10$. In Section 5, we compare the radio
properties of 3C~273 to the radio loud quasar population and find it
to be very typical. In this paper: $H_{0}$=70 km~s$^{-1}$~Mpc$^{-1}$,
$\Omega_{\Lambda}=0.7$ and $\Omega_{m}=0.3$.

\begin{figure}
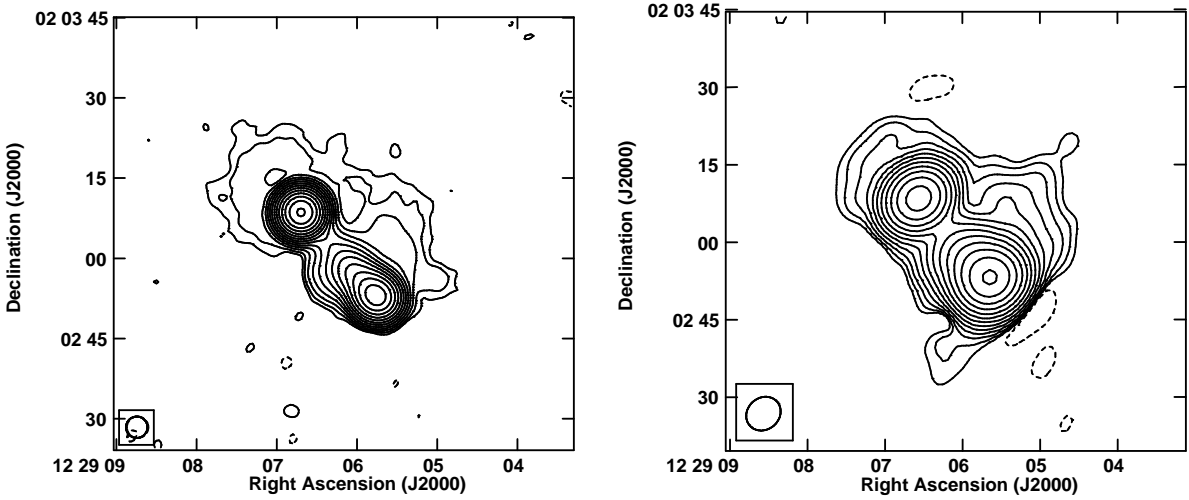

\includegraphics[width=8cm]{f1.ps}
\includegraphics[width=8cm]{f2.ps}
\caption{On the left hand sided is 3C~273 at 1365 MHz. The right
hand side is 3C~273 at 327.5 MHz. Contours in both images are in
percentage of the peak surface brightness and increase in steps of
2. The peak surface brightness and lowest contour levels are 39.7
Jy~beam$^{-1}$, $\pm0.021\%$ for the P-band image, and 32.8
Jy~beam$^{-1}$, $\pm0.011\%$ for the L-band image.}
\end{figure}

\section{Observations}
The L-band observations were carried out with the AB-array of the
VLA on July 9, 1995 (Project ID: AR334), while the P-band
observations were carried out with the VLA A-array on March 7, 1998
(Project ID: AK461). The L-band data was provided by R. Perley
\citep{per16}. The total on-source time was $\sim$30 mins in the
L-band and $\sim$1.5~hours in the P-band. We reduced and analyzed
the data using standard procedures in AIPS\footnote{Astronomical
Image Processing System}. The final {\it rms} noise in the 1365~MHz
image was $\sim1.2$~mJy~beam$^{-1}$ and the 327.5~MHz image was
$\sim1.1$~mJy~beam$^{-1}$. The baseline-based calibration task BLCAL
was used to reduce the noise in the north-south direction, which
arose due to the source being close to the equator. A dynamic range
of $\sim30,000$ and $\sim40,000$ was finally achieved at the L- and
P-bands, respectively. The radio  images with a beam-size of
$4''\times4''$ at the L-band and $7''\times6''$ at the P-band, are
presented in Figure~1. We estimated the total and extended diffuse
flux densities using the AIPS procedure TVSTAT. These were
respectively, 50.3 Jy and $0.35\pm0.04$~Jy at the L-band, and 63.3
Jy and $1.33\pm0.13$~Jy at the P-band. Defining the radio spectral
index as $F_{\nu}\propto\nu^{-\alpha}$ yields $\alpha_{327.5}^{1365}
= 0.93$.

\section{Estimating the Long Term Time Averaged Jet Power}
A method that allows one to convert 151 MHz flux densities,
$F_{151}$ (measured in Jy), into estimates of long term time
averaged jet power, $\overline{Q}$, (measured in ~ergs~s$^{-1}$) is captured
by the formula derived in \citet{wil99,pun05}:
\begin{eqnarray}
 && \overline{Q} \approx [(\mathrm{\textbf{f}}/15)^{3/2}]1.1\times
10^{45}\left[X^{1+\alpha}Z^{2}F_{151}\right]^{0.857}\mathrm{~ergs~s^{-1}}\;,\\
&& Z \equiv 3.31-(3.65)\times\nonumber \\
&&\left[X^{4}-0.203X^{3}+0.749X^{2} +0.444X+0.205\right]^{-0.125}\;,
\end{eqnarray}
where $X\equiv 1+z$, $F_{151}$ is the total optically thin flux
density from the lobes.
\par In practice, the qualifying statement that the $F_{151}$ is the total optically thin flux
density from the lobes requires a detailed study for blazars such as
3C~273. Due to Doppler boosting on kpc scales, core dominated
sources with a very bright one sided jet (such as 3C 279 and most
blazars) must be treated with care \citep{pun95}. Blazars
with significant emission on super-galactic scales (scales larger
than the host galaxy, i.e. $> 20$ kpc) typically have resolved flux
that is dominated by a one-sided jet that can be predominantly a hot
spot or strong knot. The best studied example in that paper was 3C
279, in which virtually all of the extended flux was in a one-sided
kpc jet. It was concluded that the jet dominated one sided kpc
structure was a result of strong Doppler beaming on kpc scales in
blazars. Thus, the contributions from Doppler boosted jets as well
as the radio cores must be removed before applying Equation (1).
Previous to this study, this could not be done for 3C~273. Hence,
the importance of Figure 1 and the halo flux densities derived from
them in Section 2 for estimating $\overline{Q}$.

\par The calculation of the jet kinetic luminosity in
Equation (1) depends on an empirical multiplicative factor,
\textbf{f}, that incorporates the uncertainty that is associated
with departures from minimum energy and variations in geometric
effects, filling factors, protonic contributions and the low
frequency cutoff \citep{wil99}. The quantity, \textbf{f}, was
further determined to most likely  be in the range of 10 to 20,
hence the fiducial value of 15 in Equation (1) \citep{blu00}. The
formula is most accurate for large classical double radio sources,
thus it is not applicable for sources with a linear size of less
than 20 kpc which are constrained by the ambient pressure of the
host galaxy. The halo size estimated from Figure 1 is 80 kpc by 120
kpc and likely represents both radio lobes, one on the jetted side
(south of the core) and the other on the counter jet (un-jetted)
side (north of the core). Thus, these super-galactic ``lobes" seemed
to be relaxed sufficiently in order to be considered consistent with
Equation (1). Due to the large asymmetry on the jetted and un-jetted
(counter jet) sides of the core, the kpc jet and hot spot must be
considered to be strongly Doppler boosted. Figure 1 indicates that
the Doppler enhancement of the jet and hot spot is at least three
orders of magnitude, if we assume approximate intrinsic bilateral
symmetry in the jet production. Therefore, these are not included in
determination of $F_{151}$.

\par Alternatively, one can also use the independently derived
isotropic estimator in which the lobe energy is primarily inertial
(i.e., thermal, turbulent and kinetic energy) in form \citep{pun05}
\begin{eqnarray}
&&\overline{Q}\approx
5.7\times10^{44}(1+z)^{1+\alpha}Z^{2}F_{151}\,\mathrm{~ergs~s^{-1}}\;.
\end{eqnarray}
The motivation for this derivation was the X-ray data
presented in \citet{pun05} and references therein. The data
indicates that the energy stored in radio lobes is typically
dominated by inertial energy, not magnetic field energy, contrary to
the hot spots which are often near equipartition. The derivation
then finds consistency with the spectral ageing estimates of
Fanaroff-Riley type II (FRII) radio lobes \citep{liu92}. Equation
(3) generally estimates $\overline{Q}$ lower than Equation(1) and is
considerably less for weaker radio sources such as 3C~273. Thusly
motivated, we use Equation (1) with $\mathrm{\textbf{f}} = 20$ as
the maximum upper bound on $\overline{Q}$ and Equation (3) is the
lower bound $\overline{Q}$ in the following.

\par Using $\alpha_{327.5}^{1365} = 0.93$ and the flux densities from Section 2, we estimate $F_{151}
\approx 2.7\rm{Jy}$, for the detected halo emission. However, it is
clear from Figure 1 that due to the pole-on nature of the line of
sight, a major fraction of the lobe emission projected on the sky
plane is coincident with the much brighter core, jet and hot spot.
Thus, it is not possible to extricate the halo flux from these
bright features in the overlap region. We estimate that the flux
density that is ``hidden" by the bright glow of these features is
between 50\% and 100\% of the detected extended structure. By
accounting for this hidden halo flux, we are also compensating for
the de-boosted hot spot flux that would exist if 3C~273 lay in the
sky plane, the geometric configuration for which Equations (1) and
(3) are most accurate. Thus, a more accurate estimate of the
extended flux in the lobes is

\begin{equation}
4.1\, \rm{Jy} < F_{151} < 5.5 \, \rm{Jy}\,.
\end{equation}
Applying this estimate to Equations (1) - (3) yields
\begin{equation}
7.2 \times 10^{43} \rm{~ergs~s^{-1}} < \overline{Q} < 3.7 \times 10^{44}
\rm{~ergs~s^{-1}}.
\end{equation}

\par In conclusion, the fact that the spectral index of the lobe
emission is very steep, $\alpha_{327.5}^{1365} = 0.93$, and is
distributed over a region much larger than the host galaxy (80 kpc
by 120 kpc), the lobe emission is consistent with the most important
assumption of Equations (1) and (3): the radio lobes are relaxed and
are filled with synchrotron cooling plasma. Thus, after removing the
Doppler enhanced core, jet and hot spot emission, one expects that
the estimate of $\overline{Q}$ is consistent with those performed on
classical double radio sources. This would not have been true if the
halo emission had not been detected.

\section{ The Jet Doppler Factor}
The modest value of
$\overline{Q}$ in Equation (5) needs to be reconciled with the large
apparent broadband apparent luminosity of the jet. The spectral
energy distribution (SED) of the jet has two broad components,
synchrotron emission and inverse Compton emission \citep{abd10}.
Both components are variable and the combined apparent luminosity of
the jet is $L_{\rm{app}}\sim 2 - 4 \times 10^{46} \rm{~ergs~s^{-1}}$
\citep{sol08,ghi10,abd10}. This is two orders of magnitude larger
than the value of $\overline{Q}$ in Equation (5). The SED is more
than an order of magnitude more luminous in the sub-millimeter and
mid-IR compared to 15 GHz \citep{abd10,sol08}. Consequently, most of
$L_{\rm{app}}$ is emitted from the jet in regions that are
unresolved in our radio imaging. The total apparent luminosity is
Doppler enhanced relative to the intrinsic luminosity,
$L_{\rm{int}}$, by the relationship, $L_{\rm{app}}
=\delta^{4} L_{\rm{int}}$ \citep{lig75}. The Doppler factor,
$\delta$, is given in terms of $\Gamma$, the Lorentz factor of the
outflow; $\beta$, the three velocity of the outflow and the angle of
propagation to the line of sight, $\theta$;
$\delta=1/[\Gamma(1-\beta\cos{\theta})]$ \citep{lin85}. The
discrepancy between the magnitudes of $L_{\rm{app}}$ and
$\overline{Q}$ suggests an explanation in terms of large Doppler
enhancement. In order to see if the result in Equation (5) has a
relationship to the current state of jet production, we proceed to
estimate $\delta$ by a variety of methods.

\par The Doppler factor can be constrained by a brightness temperature analysis.
When the brightness temperature in the plasma rest frame obeys,
$(T_{b})_{\rm{intr}}> 10^{12}\,^{\circ}\,\mathrm{K}\;$, the
inverse Compton catastrophe occurs \citep{kel69,mar79}. In order to
explain the observed radio synchrotron jet in such sources, Doppler
boosting is customarily invoked. This can be used to constrain the
Doppler factor in two ways. The first, we will call the
``compactness brightness temperature" argument. In this case, one
measures the size of the emission region in order to estimate
$T_{b}$ in the observers frame. If it gets too compact then an
inverse Compton catastrophe can be averted with Doppler enhancement.
In \citet{mar79}, it is argued that in order to avoid the inverse
Compton catastrophe
\begin{equation}
\delta_{\rm{com}}^{1.2} > (1+z)^{1.2}\frac{(T_{b})}{
10^{12}\,^{\circ}\,\mathrm{K}}\;.
\end{equation}
 After considering the refractive
effects and substructure, it is argued that the highest frequency,
22 GHz \emph{RadioAstron} observations, yields the most reliable
estimates of $(T_{b})$: $(T_{b}) = 1.4 \times
10^{13}\,^{\circ}\,\mathrm{K}$ \citep{kov16,joh16}. Using Equation
(6), we get a constraint on the compactness Doppler factor,
$\delta_{\rm{com}} > 10.7$.
\begin{figure}
\centering
\includegraphics[width=12cm]{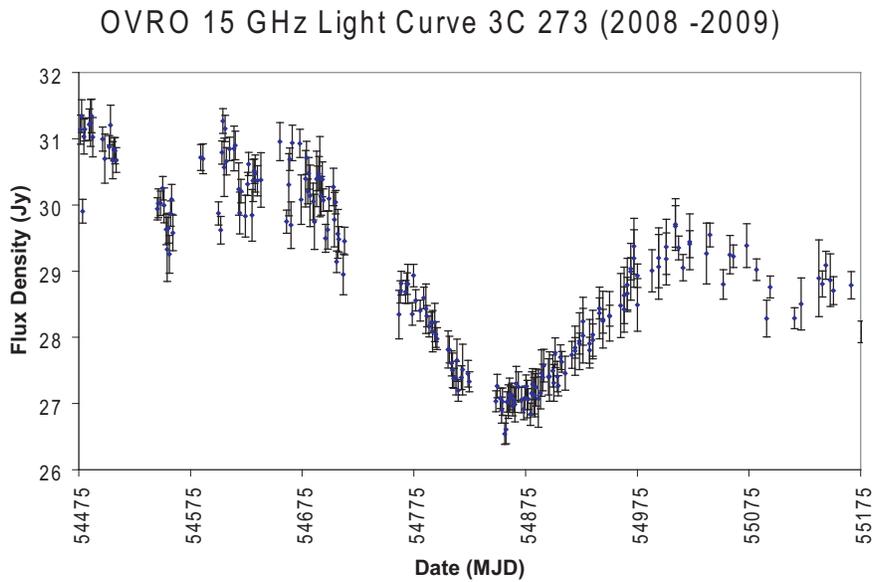}
\caption{The 15 GHz light curve from OVRO during 2008 and 2009.}
\end{figure}
\begin{figure}
\includegraphics[width=12cm]{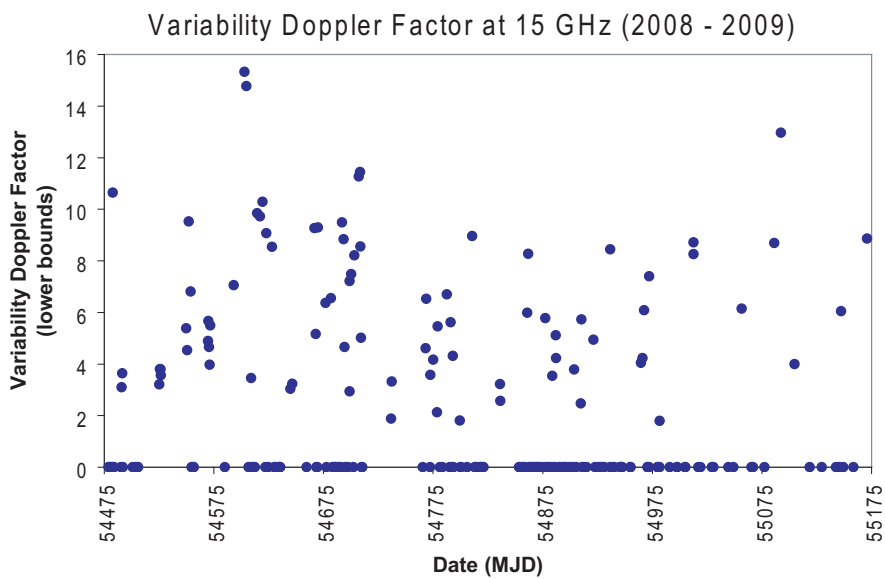}
\centering \caption{The measurements of the time variability Doppler
factor, $\delta_{\rm{var}}$, derived from the OVRO data in
Figure 2 using Equations (7) and (8). These are formally
conservative lower limits.}
\end{figure}

\par Alternatively, one can consider the constraint on $\delta$ from flux variations, ``time variability
brightness temperature" arguments. The corresponding variability
Doppler factor, $\delta_{\rm{var}}$ is defined as
\citep{gho07,hov09,zho06}
\begin{equation}
\delta_{\rm{var}} = \left[\frac{(T_{b})_{\rm{var}}}
{10^{12}\,^{\circ}\,\mathrm{K}}\right]^{1/3} \approx \left[\frac{8.0
(1+z)}{(\nu_{o}/\mathrm{1GHz})^{2}(\Delta t_{o}/ 1 \mathrm{yr})^{2}}
Z^{2} (\Delta F_{\nu}(\mathrm{mJy}))_{o}\right]^{1/3}\;,
\end{equation}
where $\Delta F_{\nu}(\mathrm{mJy})$ is the change in flux density
in mJy measured at earth at frequency, $\nu_{o}$. during the time
interval, $\Delta t_{o}$. The cosmological factor $Z$ was defined in
Equation (2). The 15 GHz OVRO (Owens Valley Radio Observatory) light
curve from 2008 to 2009 is plotted in Figure 2 \citep{ric11}. The
measurements with large errors bars ($> 0.5$ Jy) were dropped from
the plot. In order to use Equation (7), we want to be cautious of
false variability caused by measurement uncertainty. Thusly
motivated, we subtract out this uncertainty from each flux
measurement, $F_{1} \pm \sigma_{1}$ and $F_{2} \pm \sigma_{2}$ by
defining a modified flux differential
\begin{equation}
\Delta F_{\nu} \equiv \rm{maximum}\left[\left(\mid F1 - F2 \mid -
\sqrt{\sigma_{1}^{2} + \sigma_{2}^{2}}\right), \; 0 \right]\;.
\end{equation}

The time sampling by OVRO is nonuniform. Our procedure was to step
consecutively through the observations in temporal order. Each
observation, $F_{1}$ was paired with a subsequent observation,
$F_{2}$, that was as close to 10 days afterward as possible. This
procedure produced over 90\% of the 222 pairs of measurements with
separations in time from 4 to 16 days. The resulting
$\delta_{\rm{var}}$ from Equation (7) are plotted in Figure
3. It is important to note that these are lower limits. First, from
a theoretical standpoint, the inverse Compton limit is the maximum
possible brightness temperature. The system might actually exist
well below this value \citep{rea94}. Secondly, the reduced flux
differential in Equation (8) naturally produces lower values of
$\delta_{\rm{var}}$ on average in Equation (7). Thus, the
high end of the distribution is most closely related to the jet
Doppler factor. In conclusion, Figure 3 indicates that the jet
Doppler factor is above 8 and is consistent with the estimate of
$\delta_{\rm{com}} > 10.7$ found above.
\begin{figure}
\centering
\includegraphics[width=12cm]{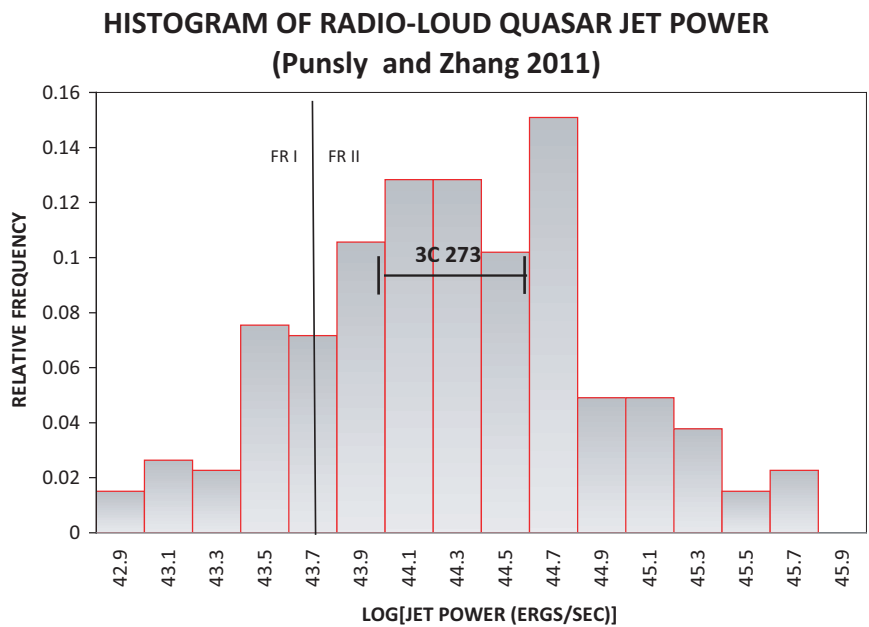}
\caption{The long term time averaged jet power of 3C~273 compared to
the distribution of jet powers for radio loud quasars. }
\end{figure}
\begin{figure}
\centering
\includegraphics[width=12cm]{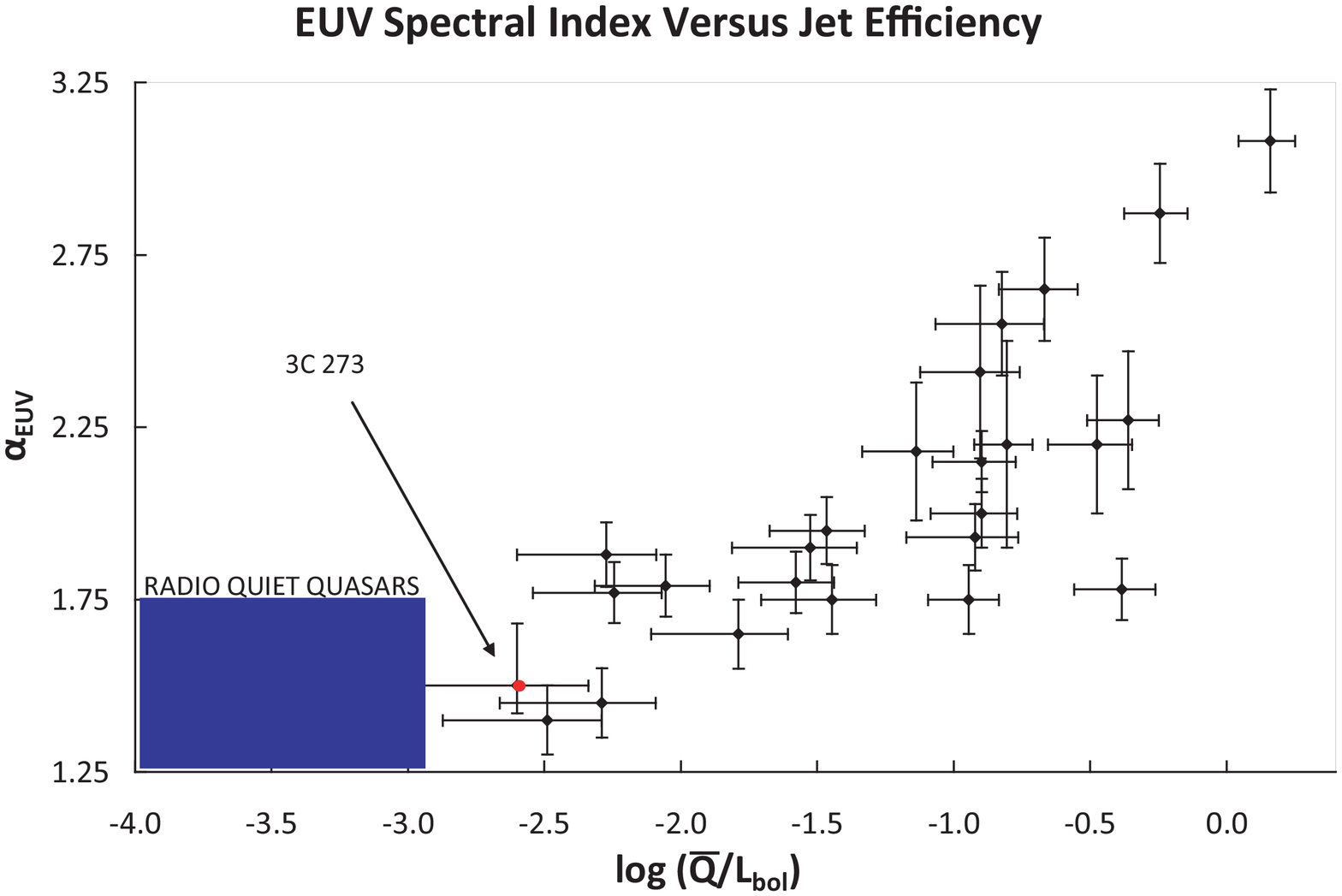}
\caption{A scatter plot of jet efficiency, $\overline{Q}/L_{bol}$,
versus $\alpha_{EUV}$ that is from \citet{pun16}. 3C~273 is
overlayed on this plot (the red dot) for comparison.}
\label{figure:20}
\end{figure}

\par Another consistency test of the larger values of $\delta \gtrsim
10$ is the large apparent velocity, $8<\beta_{\rm{app}}<15$, of
ejected components that have been monitored with the Very Long
Baseline Array \citep{lis13}. This is indicative of relativistic
motion as well. For example, at the high end, if the jet has $\Gamma
=15$ and is viewed at $3.7^{\circ}$ from the jet axis, $\delta
\approx 15.5$ and $\beta_{\rm{app}} \approx 15$. More modestly, f
the jet has $\Gamma =10$ and is viewed at $5^{\circ}$ from the jet
axis, $\delta \approx 11.5$ and $\beta_{\rm{app}}\approx 10$. Thus,
it is concluded that the preponderance of kinematical evidence
consistently indicates $\delta \gtrsim 10$.

\par Consider a large jet Doppler factor of $\delta \gtrsim 10$ in
the context of the large value of $L_{\rm{app}}\sim 2 - 4
\times 10^{46} \rm{~ergs~s^{-1}}$. Since $L_{\rm{app}} =\delta^{4}
L_{\rm{int}}$, we estimate $L_{\rm{int}}\sim 1 - 2
\times 10^{42} \rm{~ergs~s^{-1}}$. This is implies that
$L_{\rm{int}}\sim 0.01 \overline{Q}$ by Equation (5). Thus,
the large $L_{\rm{app}}$ of 3C~273 is consistent with the
Doppler enhancement of $\sim1\%$ dissipation of the time averaged
jet power.

\section{3C~273 in the Context of the Radio Loud Quasar Population}

\par The first thing to consider is how does the value of
$\overline{Q}$ for 3C~273 compare to $\overline{Q}$ of other radio
loud quasars. In Figure 4, we indicate the range of values of
$\overline{Q}$ for 3C~273 from Equation (5) relative to the
luminosity function for $\overline{Q}$ for radio loud quasars from
\citep{pun11}. The distribution of $\overline{Q}$, in Figure 4, is
from a complete sample of optically selected low redshift quasars
from the SDSS DR7 survey. The radio loud sources are all the sources
that have extended emission detected by the FIRST\footnote{Faint Images of
the Radio Sky at Twenty-centimeters \citep{Becker95}} survey on super-galactic
scales. This allows us to use the estimators in Equations (1) and
(3). The low redshift sample is pertinent since the corresponding
FIRST radio observations are sensitive enough to detect extended
flux in even the weakest FRII and many FR type I (FRI) radio
sources. Being optically selected, the sample is not skewed towards
sources with large radio flux densities. Figure 4 indicates that 3C~273 is
typical of most radio loud quasars. The $\overline{Q}$ estimates
straddle the broad maximum of the luminosity distribution.
\par Next, consider the jet power in the context of $L_{bol}$, the bolometric luminosity of the thermal emission from the accretion flow.
From \citet{pun15}, the luminosity near the peak of the spectral
energy distribution at $\lambda_{e} = 1100$\AA\ (quasar rest frame
wavelength), provides a robust estimator of $L_{\mathrm{bol}}$,

\begin{equation}
L_{bol} \approx 3.8 F_{\lambda_{e}}(\lambda_{e} = 1100 \AA) \approx
8.9 \times 10^{46}\,\mathrm{~ergs~s^{-1}}\;,
\end{equation}

where the flux density is from \citet{sha05}. Note that this
estimator does not include reprocessed radiation in the infrared
from distant molecular clouds \citep{dav11}. There exists a strong
correlation between $\overline{Q}/L_{bol}$  and the EUV (extreme
ultraviolet) spectral index that has been recently demonstrated in
radio loud quasars in a series of articles (see \citet{pun16} and
references therein). In particular, $\overline{Q}/L_{bol}$ , is
correlated with the spectral index in the EUV, $\alpha_{EUV}$;
defined in terms of the flux density by $F_{\nu} \sim
\nu^{-\alpha_{EUV}}$ computed between 700\AA\, and 1100\AA\,. The
straightforward implication is that the EUV emitting region is
related to the jet launching region in quasars. The EUV is the
highest energy optically thick emission and likely arises near the
inner edge of the accretion disk \citep{sun89,szu96}. In order to
find the location of 3C~273 in the $\overline{Q}/L_{bol}$ -
$\alpha_{EUV}$ scatter plane, we first estimate $\alpha_{EUV}$ from
the simultaneous FUSE (Far Ultraviolet Spectroscopic Explorer) and
HST (Hubble Space Telescope) observations \citep{sha05}. The
spectrum only goes down to 790\AA\, in the quasar rest frame, they
fitted the continuum from 1100\AA\, - 790\AA\, with $\alpha=
(1.34)^{+0.24}_{-0.11}$. The FUSE data needs to be extrapolated to
700\AA\, in order to compute $\alpha_{EUV}$. This is viable since
the region $800\AA\, - 900\AA\,$ has few if any broad emission lines
in quasar spectra \citep{tel02,ste14,pun15}. We estimated $\alpha=
1.9\pm 0.1$ fitted to this restricted continuum
\footnote{\citet{sco04} estimated $\alpha= 1.6\pm 0.03$ for the
entire FUSE spectral range, $790\AA\, - 1020\AA\,$.}. The $800\AA\,$
flux density is extrapolated to $700\AA\,$ with this spectral index
in order to find $\alpha_{EUV}= (1.5)^{+0.18}_{-0.08}$.  Combined
with Equations (5) and (9) this yields the placement of 3C~273 in
$\overline{Q}/L_{bol}$ - $\alpha_{EUV}$ scatter plane that was given
in \citet{pun16} and the results are shown in Figure 5. Note that 3C
273 obeys the correlation and its location in the scatter plane is
typical of quasars in which most of the energy budget is dissipated
as thermal emission and a relatively small fraction as a
relativistic jet.

\section{Conclusions}
In this article, we provided an analysis of the extended emission of
3C~273 and its implications. We presented the following results
\begin{enumerate}
\item Radio lobes are detected for the first time in the ``naked jet" quasar, 3C273.
We determine the extended halo flux densities at 1365 MHz and 327.5
MHz, to be $0.35\pm 0.04$ Jy and $1.33 \pm 0.13$ Jy, respectively.
\item We provide the first isotropic estimator of
jet power in 3C~273. Using the halo flux, we estimate a long term
time averaged jet power of $7.3 \times 10^{43} \rm{~ergs~s^{-1}} <
\overline{Q} < 3.7 \times 10^{44} \rm{~ergs~s^{-1}}$. This straddles the
peak of the radio loud quasar luminosity distribution (Figure 4)
\item It is estimated, from compactness arguments with
\emph{RadioAstron} and time variability arguments, that the Doppler
factor in the base of the jet is $\delta \gtrsim 10$, consistent
with the observations of superluminal apparent motion of components
$\sim 8c -15c$.
\item We use this estimate of the Doppler factor to constrain the
intrinsic jet broadband (radio to gamma ray) luminosity,
$L_{\rm{int}}\sim 1 - 2 \times 10^{42} \rm{~ergs~s^{-1}}$.
\item The location of 3C~273 in the
$\overline{Q}/L_{bol}$ - $\alpha_{EUV}$ plane is typical of quasars
in which most of the energy budget is dissipated as thermal emission
and a relatively small fraction as a relativistic jet (Figure 5).

\end{enumerate}

\par It seems that the estimates in points 2 and 4 are
 compatible. The implication is that if the current jet power, $Q(t)
 \sim \overline {Q}$ then $\sim 1\%$ of the jet power is dissipated as
 radiation losses.  This is a rather modest amount of dissipated
 power considering the propensity for constrained high velocity magnetized plasmas to generate
 dissipative instabilities and produce shocks; especially if the
 solar wind is an example \citep{vas03,cra09}. From points 2, 3 and 5,
 3C~273 is a prototypical radio loud quasar with an extremely large
 Doppler enhancement due to relativistic line of sight affects.

\begin{acknowledgements} We thank R. Perley for sharing his expertise and data. The National Radio Astronomy Observatory is a facility of the
National Science Foundation operated under cooperative agreement by
Associated Universities, Inc.
\end{acknowledgements}

\end{document}